\documentclass{article}
\def\p#1{\pi_{#1}}
\def\join#1#2{#1\Join#2}
\def\hoare#1#2#3{\{#1\}#2\{#3\}}
\def\crflx#1{\mean{#1}}
\def\comp{\mathbin{\cdot}}
\def\conv#1{#1^\circ}
\def\entails{\mathbin{\Rightarrow}}
\def\entailed{\mathbin{\Leftarrow}}
\def\implied{\mathbin\Leftarrow}
\def\aspas#1{``#1''}
\def\wider#1{~ #1 ~}
\def\enset#1{\mathopen{ \{ }#1\mathclose{ \} }} 
\def\deffVdm{\mathbin{\underbar\triangle}}

\let\htmladdnormallink=\quoteId		    
\usepackage{url}
\usepackage{amssymb,amsmath}
\newtheorem{definition}{Definition}
\renewcommand{\deffVdm}{\triangleq}
\usepackage{natbib}

\bibpunct[,]{(}{)}{;}{a}{,}{,}
\def\LNCS{LNCS\index{LNCS!\uk{Lecture Notes in Computer Science}}}
\usepackage[all]{xy}

\def\rarrow#1#2#3{\xymatrix{ #1 \ar[r]^-{#2} & #3 }}
\def\larrow#1#2#3{\xymatrix{ #3 & #1 \ar[l]_-{#2} }}
\let\fdep=\larrow

\let\rhoare=\rarrow
\let\longlfdep=\longlarrow
\def\lfdep#1#2#3{\larrow{#1}{#2}{#3}}
\usepackage{isolatin1}
\catcode170=13  \def^^aa{${\kern-.1em}^{\mbox{\scriptsize a}}$}      
\catcode172=13  \def^^ac{{\neg}}      
\catcode174=13  \def^^ae{\copyright{}}%
\catcode174=13  \def^^ae{\mathbin\leftarrow}
\catcode176=13  \def^^b0{^\circ}      
\catcode178=13  \def^^b2{{}^2}      
\catcode181=13  \def^^b5{\mu}				
\catcode183=13  \def^^b7{\comp}     
\catcode186=13  \def^^ba{${\kern-.1em}^{\mbox{\scriptsize o}}$}      
\catcode215=13  \def^^d7{\times}

\def\Movies{\mathit{Movies}}
\def\qemph#1{\emph{``#1''}}

\def\xarrayin#1{\begin{array}{cccccc}#1\end{array}}
\newcommand{\rcbforall}[3]{\def\nothing{}\def\range{#2}\forall\ #1 : \ifx\range\nothing #3\else \ #2\ \entails\ #3 \fi}
\newcommand{\rcbexists}[3]{\def\nothing{}\def\range{#2}\exists\ #1 : \ifx\range\nothing #3\else \ #2\ \mathbin{\wedge\,} #3 \fi}

\def\unary#1#2{\def\arg{#2}\def\omisso{}\ifx\arg\omisso{\textsf{#1}}\else{\textsf{#1}}\ {#2}\fi}
\def\ap#1#2{#1(#2)}

\def\just#1#2{\\ &#1& \rule{2em}{0pt} \{ \mbox{\rule[-.7em]{0pt}{1.8em} \small #2 \/} \} \nonumber\\ && }

\def\land{\mathbin{\ \wedge\ }}

\def\bang{{!}}

\def\ker#1{\unary{ker}{#1}}  

\def\mean#1{\mathopen{[\![}#1\mathclose{]\!]}}
\def\arrayin#1{\begin{array}{rcl}#1\end{array}}

\def\journal#1#2{\def\omissa{}\def\ref{#2}\underline{#1}\ifx\ref\omissa{}\relax\else~#2:\fi}
\def\cacm#1{\journal{CACM}{#1}}

\title{
	Functions as types or the ``Hoare logic'' of functional dependencies
}
\def\email#1{(\texttt{#1})}
\date{May 2012}
\author{Jos\'e N.\ Oliveira \\
	~\\ High Assurance Software Lab \\ INESC TEC and University of Minho \\
	4710-070 Braga, Portugal \\ \email{jno@di.uminho.pt}}

\begin{document}

\maketitle

\begin{abstract}
Inspired by the trend on unifying theories of programming, this paper shows
how the algebraic treatment of standard data dependency theory equips relational
data with \emph{functional types} and an associated type system which is
useful for type checking database operations and for query optimization.

Such a \emph{typed} approach to database programming is then shown to
be of the same family as other programming logics such as eg.\ \emph{Hoare logic}
or that of \emph{strongest invariant functions} which has been used
in the analysis of while statements.

The prospect of using automated deduction systems such as Prover9 for type-checking
and query optimization on top of such an algebraic approach is considered.
\end{abstract}

\section{Prelude}
\label{sec:120305a}
In a paper addressing the influence of Alfred Tarski (1901-83) in computer
science, Solomon \citet{Fe06} quotes the following statement by his
colleague John Etchemendy:
\qemph{%
	You see those big shiny Oracle towers on Highway 101? They would never
have been built without Tarski's work on the recursive definitions of satisfaction
and truth}.

The `big shiny Oracle towers' are nothing but the headquarters of Oracle
Corporation, the giant database software provider sited in the San Francisco
Peninsula. Still \citet{Fe06}:
\qemph{%
Does Larry Ellison know who Tarski is or anything about his work?
\em [...] \em
I learned subsequently from Jan Van den Bussche that
\em [...] \em
he marks the reading of Codd's seminal paper as the 
starting point leading to the Oracle Corporation.}

\citet{Bu01} had in fact devoted attention to relating Codd and Tarski's work:
\qemph{%
We conclude that Tarksi produced two alternatives for Codd's relational algebra:
{cylindric} set algebra, and relational algebra with {pairing} {\em[...]}
For example, we can represent the ternary relation
$\enset{(a,b,c),(d,e,f)}$ as
$\enset{(a,(b,c)),(d,(e,f))}$}.
Still \cite{Bu01}:
\begin{quote}\qemph{%
Using such representations, we leave it as an exercise to the reader to
simulate Codd's relational algebra in RA$^+$ {\em[relational algebra with pairing]}}.
\end{quote}

To the best of the author's knowledge, nobody has thus far addressed this \emph{exercise}
in a thorough way. Instead, standard relational database theory \citep{Ma83,AHV95} includes a
well-known relation algebra but this is worked out in set theory and quantified
logic, far from the objectives of Tarski's life-long pursuit in developing
methods for elimination of quantifiers from logic expressions. An effort
which ultimately lead to his \emph{formalization of set theory without variables}
\citep{TG87}. 

The topic has acquired recent interest with the advent of work on
implementing extensions of Tarski's algebra in automated deduction systems
such as \emph{Prover9} and the associated counterexample generator \emph{Mace4} \citep{HS07}.
This offers a potential for automation which has not been acknowledged
by the database community.
In this context, it is worth mentioning an early concern of the founding
fathers of the standard theory \citep{BRH77}:
\begin{quote}\qemph{%
	[A] {general theory} that ties together {dependencies},
	{relations} and {operations} on relations is still lacking}.
\end{quote}
More than 30 years later, this concern is still justified, as database programming
standards remain insensitive to techniques such as formal verification and
\emph{extended static checking} \citep{FLLNSS02} which are more and more regarded
essential to ensuring quality in complex software systems.

In the remainder of this paper we will see how the algebraic treatment of
the standard theory along the \emph{exercise} proposed by Bussche equips
relational data with \emph{functional types} and an associated type system
which can be used to type check database operations. Interestingly,
such a \emph{typed} approach to database programming will be shown to
relate to other programming logics such as eg.\ \emph{Hoare logic} \citep{Ho69}
or that of \emph{strongest invariant functions} \citep{MDG85} which has been used
in the analysis of while statements, for instance.

On the whole, the approach has a \emph{unifying theories of programming}
\citep{HoJi98} flavour, even though the exercise is not carried out ``avant
la lettre'' in canonical UTP. A full account can be found in a technical
report \citep{Ol11}. For space constraints, this paper only covers the first
part of the exercise, that of developing a type system for relational data
which stems from functional dependencies.

\paragraph{Paper structure.} Section \ref{sec:120303c} introduces functional
dependencies (FD) and shows how to convert the standard definition into the
Tarskian, quantifier-free style. The parallel between the \emph{functions
as types} approach which emerges from such a conversion and a similar treatment
of Hoare logic is given in section \ref{sec:120312a}. Section \ref{sec:120312b}
shows that, in essence, \emph{injectivity} is what matters in FDs and gives
a corresponding, simpler definition of FD which is used in section \ref{sec:120312c} to
re-factor the standard theory into a \emph{type system of FDs}. Section
\ref{sec:120312d} shows how to use this type system to type check database
operations and section \ref{sec:120312e} shows how to calculate query optimizations
from FDs. The last section gives an account of related work and concludes with
a prospect for future work.

\section{Introducing functional dependencies}
\label{sec:120303c}
In standard relational data processing, real life objects or entities are recorded by assigning
values to their observable properties or \emph{attributes}. A database file
(vulg.\ \emph{table})
is a collection of such attribute assignments, one per object, such that
all values of a particular attribute (say $i$) are of the same type (say $A_i$).
 For $n$ such attributes, a \emph{relational database file} $T$ can be regarded
as a set of $n$-tuples, that is, $T \subseteq A_1 × \ldots × A_n$. A \emph{relational
database} is just a collection of several such $n$-ary relations, or tables.

Attribute names normally replace natural numbers in the identification of
attributes. The enumeration of all attribute names in a database table,
for instance
\(
    S \wider= \enset{\textsc{Pilot}, \textsc{Flight}, \textsc{Date}, \textsc{Departs}}
\)
concerning the airline scheduling system given as example in \citep{Ma83},
is a finite set called the table's \emph{scheme}.
This scheme captures the \emph{syntax} of the data.
What about \emph{semantics}?
Even non-experts in airline scheduling will accept ``business rules" such as,
for instance:
\begin{em}
     a single pilot is assigned to a given flight, on a given date.
\end{em}
This restriction is an example of a so-called \emph{functional dependency}
(FD) among attributes, which can be stated more formally
by writing
\qemph{\textsc{Flight} \textsc{Date} $\rightarrow$ \textsc{Pilot}}
to mean that \emph{attribute $\textsc{Pilot}$ is functionally dependent on $\textsc{Flight}$
and $\textsc{Date}$}, or that \emph{$\textsc{Flight}, \textsc{Date} $ functionally
determine $\textsc{Pilot}$}.

Data dependencies capture the \emph{meaning} of relational data.
Data dependency theory involves not only functional
dependencies (FD) but also multi-valued dependencies (MVD). Both are central
to the standard theory, where they are addressed in an axiomatic way. \cite{Ma83}
provides the following definition for FD-satisfiability: 
\begin{definition} \label{def:041216a}
Given subsets $x,y \subseteq S$ of the relation scheme $S$ of a $n$-ary relation $T$,
this relation is said to satisfy functional dependency $x\rightarrow y$
iff all pairs of tuples $t,t'\in T$ which \aspas{agree} on $x$ also \aspas{agree} on $y$,
that is,
\begin{eqnarray}
	\rcbforall{t,t'}{t,t'\in T}{(\arrayin{t[x]=t'[x] &\entails& t[y] = t'[y] })}
	\label{eq:041214a}
\end{eqnarray}
(The notation $t[a]$ in (\ref{eq:041214a}) means \aspas{the value exhibited
by attribute $a$ in tuple $t$}.)
\\ $\Box$
\end{definition}

How does one express formula (\ref{eq:041214a}) in Tarski's relation algebra
style, getting way with the two-dimensional universal quantification and 
logical implications inside?
For so doing we need to settle some notation.
To begin with, $t[x]$
is better written as $\ap x t$, where
$x$ is identified with the \emph{projection function} associated to attribute
set $x$. Regarding $x$ and $y$ in (\ref{eq:041214a}) as such functions we write:
\begin{eqnarray}
	\rcbforall{t,t'}{t,t'\in T}{(\arrayin{\ap x t=\ap x{t'} &\entails& \ap{y}{t} = \ap{y}{t'} })}
	\label{eq:120302a}
\end{eqnarray}

Next, we observe that, given a function $f: A \rightarrow B$,
the binary relation $R \subseteq A \times A$
which checks whether two values of $A$
have the same image under $f$~\footnote{This is known as the \emph{nucleous}
	\citep{MDG85} or \emph{kernel} \citep{Ol08b} of a function $f$.}
--- that is,
$a' R a \equiv {\ap{f}{a'}}={\ap{f}{a}}$ ---
can be written alternatively as $a'(\conv f \comp f)a$.
Here, $\conv f$ denotes the \emph{converse}
of $f$ (that is, $a(\conv f)b$ holds iff $b=f\ a$) and the dot ($\comp$)
denotes the extension of function composition to binary relations:
\begin{eqnarray}
	b (R·S) c & \wider\equiv & \rcbexists a{}{b \ R \ a \land a \ S \ c}
	\label{eq:051118b-def}
\end{eqnarray}

Using converse and composition the rightmost implication of (\ref{eq:120302a})
can be rewritten into
\(
	t(x° · x) t'
	\entails
	t(y° · y) t'
\),
for all $t,t'\in T$.
Implications such as this can expressed as relation inclusions, following 
definition:
\begin{eqnarray}
	R \subseteq S & \wider\equiv & \rcbforall{b,a}{}{b\ R\ a \entails b\ S\ a}
	\label{eq:041216a}
\end{eqnarray}
However, just stating the inclusion $x° · x \subseteq y° · y$ would be a gross
error, for the double scope of the quantification ($t\in T \land t'\in T$) would
not be taken into account.
To handle this, we first unnest the two implications of (\ref{eq:120302a}),
\begin{eqnarray*}
	\rcbforall{t,t'}{}{(t\in T \land t'\in T \land t(x° · x) t')
	\entails
	t(y° · y) t'}
\end{eqnarray*}
and treat the antecedent $t\in T \land t'\in T \land t(x° · x) t'$ independently,
by replacing the set of tuples $T$ by the binary relation $\crflx T$ defined as follows
\footnote{
This is a standard way of encoding a set $T$ as a binary relation $\crflx T$
known as a \emph{partial identity}, since $\crflx T \subseteq id$. The set
of all such relations forms a Boolean algebra which reproduces the usual algebra
of sets. Moreover, partial identities are symmetric ($\conv{\crflx T}=\crflx T$) and
such that
$\crflx S \comp \crflx T = \crflx S \cap \crflx T$.}:
\begin{eqnarray}
	b \crflx T a & \wider\equiv & b=a \land a\in T
	\label{eq:120302b}
\end{eqnarray}
Note that $t\in T$ can be expressed in terms of $\crflx T$ by
$\rcbexists{u}{u=t}{t \mean T u}$ and similarly for $t'\in T$.
Then:
\begin{eqnarray*}
&&
	(t\in T \land t'\in T \land t(x° · x) t')
\just\equiv{ expansion of $t\in T$ and $t'\in T$ }
	\rcbexists{u,u'}{}{
	u = t 
	 \land 
	u' = t'
	\land
	t \mean T u
	 \land 
	t' \mean T u'
	 \land 
	t(x° · x) t'
	}
\just\equiv{$\land$ is commutative; equal by equal substitution; converse }
	\rcbexists{u,u'}{}{
	t \mean T u  \land u(x° · x) u' \land u' \conv{\mean T} t'
	}
\just\equiv{ composition (\ref{eq:051118b-def}) twice  }
	t(\mean T \comp x° · x \comp \conv{\mean T}) t'
\end{eqnarray*}
Finally, by putting this together with $t(y° · y)t'$ we obtain
\begin{eqnarray}
	\mean T \comp x° · x \comp \conv{\mean T} \subseteq y° · y
	\label{eq:120303a}
\end{eqnarray}
as a quantifier-free relation algebra expression meaning the same as
(\ref{eq:041214a}). 

\paragraph{Generalization.}
To reassure the reader worried about the doubtful practicality of derivations
such as the above, we would like to say that we don't need to do it over
and over again: inequality (\ref{eq:120303a}), our Tarskian alternative to
the original textbook definition (\ref{eq:041214a}), is all we need for calculating
with functional dependencies.  Moreover, we can start this by actually expanding
the scope of the definition from sets of tuples $\crflx T$ and attribute
functions ($x$, $y$) to arbitrary binary relations $R$ and suitably typed
functions $f$ and $g$:
\begin{eqnarray}
	R \comp f° · f \comp \conv{R} \subseteq g° · g
	\label{eq:120303b}
\end{eqnarray}

In this wider setting, $R$ can be regarded not only as a piece of data but
also as the specification of a nondeterministic computation, or even the
transition relation of a finite-state automaton; and $f$ (resp.\ $g$)
as a function which observes the input (resp.\ output) of $R$.
Put back into quantified logic, such a wider notion of a functional dependency
will expand as follows:
\begin{eqnarray}
	\rcbforall{a',a}{{\ap{f}{a'}}={\ap{f}{a}}}
		{(\rcbforall{b',b}{b'\ R\ a' \land b\ R\ a}{{\ap{g}{b'}}={\ap{g}{b}}})}
	\label{eq:120304a}
\end{eqnarray}
In words: \emph{inputs $a$, $a'$ indistinguishable by $f$ can via $R$ only lead to outputs
indistinguishable by $g$}. Notationally, we will convey this interpretation
by writing $R : f \rightarrow g$ or $\rarrow f R g$. We can still say that
$R$ satisfies the $f \rightarrow g$ FD, in particular wherever $R$ is a 
piece of data. As can be easily checked, ${\ap{f}{a'}}={\ap{f}{a}}$ is an equivalence
relation which, in the wider setting, can be regarded as the \emph{semantics}
of the datatype which $R$ takes inputs from
(think of $f: A \to B$ as a \emph{semantic} function mapping a syntactic
domain $A$ into a semantic domain $B$), and similarly for $g$ concerning
the output type.

Summing up, the functions $f$ and $g$ in (\ref{eq:120303b}) can be regarded
as \emph{types} for $R$. Some type assertions of this kind will be very easy
to check, for instance $id : f \to f$, just by replacing $R,f,g :=id,f,f$
in (\ref{eq:120303b}) and simplifying. But type inference will be easier
to calculate on top of the even simpler (re)statement of (\ref{eq:120303b})
which is given next.

\section{Functions as types}
\label{sec:120312a}
Before proceeding let us record two properties of the relational operators
\emph{converse} and \emph{composition}~\footnote{It may help to recall the same properties
from elementary linear algebra, once converse is interpreted as
matrix transposition and composition as matrix-matrix multiplication.}:
\begin{eqnarray}
	(R \comp S)° &=& S° \comp R°
	\label{eq:020624d}
\\
	(R°)° &=& R
	\label{eq:020624b}
\end{eqnarray}
Moreover, it will be convenient to have a name for the relation $\conv R \comp R$
which, for $R$ a function $f$, is the equivalence relation ``indistinguishable by $f$'' seen above. We define
\begin{eqnarray}
	\ker R & \wider\deffVdm & \conv R \comp R
	\label{eq:020624a}
\end{eqnarray}
and read $\ker R$ as ``the \emph{kernel} of $R$''.
Clearly, $a'(\ker R)a$ means $\rcbexists b {}{b\ R\ a' \land b\ R\ a}$ and therefore
$\ker R$ measures the \emph{injectivity} of $R$: the larger it is the larger the
set of inputs which $R$ is unable to distinguish (= the less \emph{injective} $R$ is).

We capture this by introducing a preorder on relations which compares
their \emph{injectivity}:
\begin{eqnarray}
	R \leq S & \wider\deffVdm & \ker S \subseteq \ker R
	\label{eq:041217a}
\end{eqnarray}
As an example, take two list functions, $\mathit{elems}$ computing the
set of all elements of a list, and $\mathit{bagify}$ keeping the bag of
such elements. The first loses more information (order and multiplicity)
than the latter, which only forgets about order. Thus $\mathit{elems} \leq \mathit{bagify}$.
A function $f$ (relation in general) will be \emph{injective} iff $\ker f \subseteq id$
($id \leq f$),
which easily converts to the usual definition:
\(
	{\ap{f}{a'}} = {\ap{f}{a}} \entails a' = a
\).

Summing up: for functions or any totally defined relations
$R$ and $S$~\footnote{A
relation $R$ is totally defined (or \emph{entire}) iff $id \subseteq \ker R$.},
$R\leq S$ means that \emph{$R$ is less injective than $S$};
for possibly partial $R$ and $S$, it will mean that \emph{$R$ less injective
or more defined than $S$}.

Therefore, for \emph{total} relations $R$ the preorder is universally bounded,
\begin{eqnarray*}
	\xarrayin{ \bang & \leq & R & \leq & id  }
\end{eqnarray*}
where the infimum is captured by the constant function $!$ which maps every argument
to a given (predefined) value, the choice of such value being irrelevant~\footnote{
Note that $R \leq S$ is a preorder, not a partial order, meaning that two
relations indistinguishable with respect to their degree of injectivity can be different.}.
The kernel of $\bang$ is therefore the largest possible, denoted by $\top$ (for
``top''). The other bound is trivial to check, since $\ker{id} = id$,
this arising from the well-known fact that $id$ is the unit of composition.
In general, $id \leq R$ means $R$ is injective.

Equipped with this ordering, we may spruce up our relational characterization
of the $\rarrow f R g$ type assertion, or functional dependency (FD):
\begin{eqnarray*}
&&
	\rarrow fRg
\just\equiv{ definition (\ref{eq:120303b}) }
	R \comp f° · f \comp \conv{R} \subseteq g° · g
\just\equiv{ converses (\ref{eq:020624d},\ref{eq:020624b}) ; kernel (\ref{eq:020624a}) }
	\ker{(f· R°)} \subseteq \ker g
\just\equiv{ (\ref{eq:041217a}): $g$ is ``\emph{less injective} than $f$ wrt.\ $R$'' }
	g \leq f· R°
\end{eqnarray*}
We thus reach a rather elegant formula for expressing functional dependencies,
whose layout invites us to actually swap the direction of the arrow notation (but,
of course, this is just a matter of taste):

\begin{definition}
Given an arbitrary binary relation $R \subseteq A \times B$
and functions $f : B \to D$ and $g : A \to C$, given $A$, $B$, ... $D$,
the ``type assertion'' $\larrow f R g$ meaning that
\emph{$R$ satisfies FD $f \rightarrow g$} is given by
the equivalence:
\begin{eqnarray}
	\larrow f R g ~ & \equiv & ~ g \leq f · R°
	\label{eq:050109b}
\end{eqnarray}
$\Box$
\end{definition}
Intuitively, $\larrow f R g$ means that $g$ will be \emph{blinder}
(less injective) to the outputs of $R$ than $f$ is concerning its inputs.

There are two main advantages in definition (\ref{eq:050109b}), besides saving ink.
The most important is that it takes advantage of the calculus of injectivity which
will be addressed in the following section. The other is that it makes it easy
to bridge with other programming logics, as is seen next.

\paragraph{Parallel with Hoare logic.}
As is widely known, Hoare logic is based on triples of the form $\hoare
p R q$, with the standard interpretation: \emph{``if the assertion $p$ is
true before initiation of a program $R$, then the assertion $q$ will be true
on its completion"} \citep{Ho69}.

Let program $R$ be identified with the 
relation which captures its state transition semantics and predicates $p$ (and $q$) 
be identified with $s'\crflx p s \wider\equiv s'=s \land  \ap p s$ (similarly
for $q$) --- the same trick we used for converting sets to binary relations
in section \ref{sec:120303c}. (Note how $\crflx p$ can be regarded as 
the semantics of a statement which
checks $\ap p s$ and does not change state, failing otherwise.)
In relation algebra this is captured by~\footnote{See \citep{Ol08b} and references there
to related work.}
\begin{eqnarray*}
	\hoare p R q & \wider\equiv & rng(R \comp \crflx p) \subseteq \crflx q
\end{eqnarray*}
meaning that the outputs of $R$ (given by the range operator $rng$) for inputs
pre-conditioned by $p$ don't fall outside $q$; that is, $q$ is \emph{weaker}
than the strongest post-condition $sp(R,p)$, something we can express by
writing
\begin{eqnarray}
	\hoare p R q & \wider\equiv & q \leq p \comp \conv R
	\label{eq:120303e}
\end{eqnarray}
under a suitable preorder $\leq$ expressing that $q$ is less constrained than
$p \comp \conv R$~\footnote{Details: $\hoare p R q$ is
$rng(R \comp \crflx p) \subseteq \crflx q$,
itself the same as $dom(\crflx p \comp \conv R) \subseteq dom \crflx q$
since $dom$ (domain) and $rng$ (range) commute with converse and
the domain of a partial identity is itself. The preorder is
$R \leq S \equiv dom\ S \subseteq dom\ R$. Parentheses $\crflx \_ $ are dropped
to make the formula lighter to read.
}. 

In spite of the different semantic context, there is a striking formal similarity
between formulas (\ref{eq:120303e}) and (\ref{eq:050109b}) suggesting that
Hoare logic and the logic we want to build for FDs share the same mathematics
once expressed in relation algebra. Such similarities will become apparent
in the sequel, where we are going to write $\rhoare p R q$ for $\hoare p R q$, to
put the notations closer. Using this notation, rules such as eg. the rule
of composition,
\(
	\hoare p {R_1} q \land \hoare q {R_2} r
	\entails
	\hoare p {R_1 ; R_2} r
\) become \footnote{
The arrow notation for Hoare triples, reminiscent of that
of labelled transition systems, is adopted in eg.\ \citep{Ol08b}.}:
\begin{eqnarray}
	\rhoare{p}{R_1}{q} \land \rhoare{q}{R_2}{r}
	& \entails &
	\rhoare p{R_1 ; R_2}r
	\label{eq:120303d}
\end{eqnarray}
We will check the FD equivalent to (\ref{eq:120303d}) shortly. 

\section{A calculus of injectivity ($\leq$)}
\label{sec:120312b}
One of the advantages of relation algebra is its easy ``tuning'' to
special needs, which we will illustrate below concerning the
algebra of injectivity. We give just an example,
taken from \citep{Ol11}; the reader is referred to this technical
report for the whole story.

We start by considering two rules of relation algebra which prove
very useful in program calculation:
\begin{eqnarray}
	f · R \subseteq S & \equiv & R  \subseteq f ° · S
	\label{eq:020617e}
\\
	R · f° \subseteq S & \equiv & R  \subseteq S · f
	\label{eq:020617f}
\end{eqnarray}
In these equivalences~\footnote{Technically, these equivalences should be
regarded as (families of) Galois connections \citep{Ol08b}.}, which are widely  known as
\emph{shunting rules} \citep{BM97},
$f$ is required to be a (total) function. In essence, they let one trade 
a function $f$ from one side to the other of a $\subseteq$-equation
just by taking converses. (This is akin to ``changing sign'' in trading terms in
inequations of elementary algebra.)

It would be useful to have similar rules for the injectivity preorder,
which we have chosen as support for our definition of a FD
(\ref{eq:050109b}).
It turns out that such rules are quite easy to infer,
as is the case of the Galois connection for trading a function $f$
with respect to the injectivity preorder given by
\begin{eqnarray}
	R · f \leq S & \wider\equiv & R \leq S · { f° }
	\label{eq:050118a}
\end{eqnarray}
which takes just three steps to calculate:
\begin{eqnarray*}
&&
	R · f \leq S
\just\equiv{ definition (\ref{eq:041217a}) ; converses (\ref{eq:020624d},\ref{eq:020624b}) ; kernel (\ref{eq:020624a}) }
	\ker S \subseteq f°· (\ker R) · f
\just\equiv{ shunting rules (\ref{eq:020617e},\ref{eq:020617f}) }
	f· \ker S· f° \subseteq \ker R
\just\equiv{ converses, kernel and definition (\ref{eq:041217a}) again }
	R \leq S· f°
\end{eqnarray*}

Let us put this new rule to work for us in the
derivation of a \emph{trading}-rule which will enable handling
composite antecedent and consequent functions in FDs:
\begin{eqnarray}
	\lfdep x {z· R· k°} y & \equiv & \lfdep {x· k} {R} {y· z}
	\label{eq:061227a-modified}
\end{eqnarray}
Thanks to (\ref{eq:050118a}), the calculation of (\ref{eq:061227a-modified}) is immediate:
\begin{eqnarray*}
&&
	\lfdep x {z· R· k°} y
\just\equiv{ definition (\ref{eq:050109b}) ; converses }
	y \leq x· k· R°· z°
\just\equiv{ new shunting rule (\ref{eq:050118a}) }
	y· z \leq (x· k)· R°
\just\equiv{ definition (\ref{eq:050109b}) }
	\lfdep {x· k} {R} {y· z}
\end{eqnarray*}

Another result which will help in the sequel is
\begin{eqnarray}
	X \leq R \cup S & \wider\equiv& X \leq R \wider\land X \leq S \wider\land R°· S \subseteq \ker X
	\label{eq:091201c}
\end{eqnarray}
where $R \cup S$ is the union of relations $R$ and $S$~\footnote{%
See \citep{Ol11} for the proof of this and
other results of the algebra of injectivity.}.
For
	 $X := id$, (\ref{eq:091201c}) tells that
	 $R \cup S$ is injective \textbf{iff} both
	 $R$ and
	 $S$ are injective and \emph{don't ``confuse'' each other}:
wherever
	 $b S a$ and
	 $b R c$ hold,
	 $c=a$.

\section{Building a type system of FDs}
\label{sec:120312c}
The machinery set up in the previous sections is enough for developing a
type system whereby \emph{dependencies, relations and operations on relations
are tied together}, as \cite{BRH77} envisaged.

\paragraph{Composition rule.} FDs on relations which matching antecedent and
consequent functions (as types) compose:
\begin{eqnarray}
	\lfdep {x} {S · R} {y} & \implied &
	\arrayin{ \lfdep{z}{S}{y} & \land & \lfdep{x}{R}{z} }
	\label{eq:041106a}
\end{eqnarray}
\emph{Proof:}
\begin{eqnarray*}
&&
\arrayin{
	\lfdep g S h
	& \land &
	\lfdep f R g
	}
\just\equiv{ (\ref{eq:050109b}) twice }
\arrayin{
	h \leq g · S°
	& \land &
	g \leq f · R°
	}
\just\entails{ $\leq$-monotonicity of $(~ · S°)$ 
	; converse (\ref{eq:020624d})
	}
\arrayin{
	h \leq g · S°
	& \land &
	g · S° \leq f · (S · R)°
	}
\just\entails{ $\leq$-transitivity }
\arrayin{
	h \leq f · (S · R)°
	}
%
\just\equiv{ (\ref{eq:050109b}) again }
	\lfdep f {S · R} h
\end{eqnarray*}

This rule is the FD counterpart of the rule of composition in Hoare logic (\ref{eq:120303d})
for $R$ and $S$ regarded as describing computations~\footnote{
For $R$ and $S$ the same database table, this rule subsumes Armstrong axiom F5 (Transitivity)
in the standard FD theory \citep{Ma83}. The calculation of this and other similar
results stated in this paper can be found in \citep{Ol11}.}.

\paragraph{Consequence (weakening/strengthening) rule:}
\begin{eqnarray}
	\xarrayin{\lfdep {h} {R} {k}} & \implied & \xarrayin{ k \leq g & \land & \lfdep {f} {R} {g} & \land & f \leq h }
	\label{eq:041118a}
\end{eqnarray}
\emph{Proof:} See \citep{Ol11}, where this rule is shown to
subsume and generalize standard Armstrong axioms F2 \emph{(Augmentation)} and F4 \emph{(Projectivity)}.
In the parallel with Hoare logic, it corresponds to the two \emph{rules of consequence}
\citep{Ho69} which, put together and writing triples as arrows, becomes
\begin{eqnarray*}
	\lfdep {p'} {P} {q'} & \implied & \xarrayin{ {q'} \implied q & \land & \lfdep {p} {P} {q} & \land & p \implied {p'} }
\end{eqnarray*}
for $P$ a program and $p$, $q$ etc program assertions.

\paragraph{Reflexivity.}
We have seen already that
\begin{eqnarray}
	\lfdep f {id} f
	\label{eq:041106c}
\end{eqnarray}
holds trivially. 
This rule, which corresponds to the ``\emph{skip}'' rule of Hoare logic,
\(
	\lfdep p {skip} p
\),
is easily shown to hold for any set $T$,
\begin{eqnarray}
	\lfdep f {\crflx T} f
	\label{eq:041118d}
\end{eqnarray}
as FDs are downward closed.
Rule (\ref{eq:041118d}) is known as Armstrong axiom F1 (Reflexivity).

Note in passing that (\ref{eq:041106a}) and (\ref{eq:041106c}) together
define a {category} whose objects are functions (types) and whose 
morphisms (arrows) are FDs.

\section{Type checking database operations}
\label{sec:120312d}
Let us proceed to an example of database operation type checking:
we want to know what it means for the merging of two database files
to satisfy a particular functional dependency $\rarrow f{}g$.
That is, we want to find a \emph{sufficient} condition for
the union $R \cup S$ of two relations $R$ and $S$ to be of type
$\rarrow f{}g$.
The algebra of injectivity does most of the work:
\begin{eqnarray*}
&&
	\lfdep f {R\cup S} g
\just\equiv{ definition (\ref{eq:050109b}) ; converse distributes by union }
	g \leq f· (R° \cup S°)
\just\equiv{ relational composition distributes through union }
	g \leq f· R° \cup f· S°
\just\equiv{ algebra of injectivity (\ref{eq:091201c}); definition (\ref{eq:050109b}) again, twice }
	\lfdep f R g \wider\land 
	\lfdep f S g \wider\land
	R· \ker f· S°\subseteq \ker g
\just\equiv{ introduce ``mutual dependency'' shorthand }
	\lfdep f R g \wider\land 
	\lfdep f S g \wider\land
	\lfdep f {R,S} g
\end{eqnarray*}

The ``mutual dependency'' shorthand $\lfdep f {R,S} g$ introduced in the
last step for $R· \ker f· S°\subseteq \ker g$ can be read as a generalization
of the standard definition of a FD to \emph{two} relations instead of \emph{one}
--- just generalize the second $R$ in (\ref{eq:120304a}) to some $S$. 
For $R$ and $S$ two sets of tuples, it means that grabbing one tuple from one
set and another tuple from the other set, if they cannot be distinguished by $f$
then they will remain indistinguishable by $g$.

It should be stressed that the bottom line of the calculation expresses not
only a \emph{sufficient} but also a \emph{necessary} condition for
$\lfdep f {R\cup S} g$ to hold,
as all steps are equivalences.

Type checking other database operations will follow the same scheme. Below we handle
one particular such operation --- relational \emph{join} \citep{Ma83} --- in detail.
This is justified not only for its relevance in data processing but also
because it brings about other standard FD rules not yet addressed.

\paragraph{Joining and pairing.}
Recall from section \ref{sec:120305a} how \cite{Bu01} explains the relevance of Tarski's
work on \emph{pairing} in relation algebra by illustrating how a ternary
(in general, $n$-ary)
relation $\enset{(a,b,c),(d,e,f)}$ gets represented by a binary one,
$\enset{(a,(b,c)),(d,(e,f))}$.

Pairing is not only useful for ensuring that sets of arbitrarily long (but
finite) tuples are represented by binary relations but also for defining
the \emph{join} operator ($\Join$) on such sets.  In fact, this operator
is particularly handy to express in case the two sets of tuples are already
represented as binary relations $R$ and $S$:
\begin{eqnarray}
	\arrayin{
	(a,b)(\join{R}{S}) c & \wider\equiv & a \ R \ c \land b \ S \ c
	}
&&
	\label{eq:030418aPW}
\end{eqnarray}
Interestingly, relational join behaves as a \emph{least upper bound} with respect to
the injectivity preorder~\footnote{See details and proof in \citep{Ol11}.}:
\begin{eqnarray}
	\join  R S \leq T & \wider\equiv &
	R \leq T \land S \leq T
	\label{eq:050112a}
\end{eqnarray}

This combinator turns out to be more general than its use in data processing~\footnote{It
is termed \emph{split} in \citep{BM97} and \emph{fork} in \citep{FBH97}.}.
In particular, when $R$ and $S$ are functions $f$ and $g$, $f\Join g$ is the obvious
function which pairs the outputs of $f$ and $g$:
\(
	(f\Join g)x \wider= (f\ x, g\ x)
\).
Think for instance of the projection function $f_x$ (resp.\ $f_y$) which, in the context of
Definition \ref{def:041216a} yields $t[x]$ (resp.\ $t[y]$) when applied to a tuple $t$.
Then $(f_x\Join f_y)t=(t[x],t[y])=t[xy]$, where $xy$ denotes the union of attributes
$x$ and $y$ \citep{Ma83}. So, attribute union corresponds to joining the
corresponding projection functions. This gives us a quite uniform framework for
handling both relational join and compound attributes. To make notation closer to what
is common in data dependency theory we will abbreviate $f_x\Join f_y$ to $f_xf_y$ and
this even further to $xy$, identifying (as we did before) each attribute (eg.\ $x$) with
the corresponding projection function (eg.\ $f_x$).

Minding this abbreviation $f g$ of $f \Join g$, for functions,
from (\ref{eq:050112a}) it is easy to derive facts
\(
	\xarrayin{ \bang & \leq & f & \leq & id  }
\)
and
\(
	\xarrayin { f \leq \      f g & \mbox{~,~} & g \leq \      f g }
\).
This is consistent with the use of juxtaposition to denote ``sets of attributes''.
In this context, $\leq$ can be regarded as expressing ``attribute inclusion''.
In fact, the more attributes one observes the more injective the projection
function corresponding to such attributes is~\footnote{This parallel between
attribute sets ordered by inclusion and projection functions ordered by
injectivity is dealt with in detail in \citep{Ol11}. Note how $\bang$ mimics
the empty set and $id$ mimics the whole set of attributes, enabling one to ``see
the whole thing'' and thus discriminating as much as possible.}.

A first illustration of this unified framework is given below: the (generic) calculation
of the so-called Armstrong axioms F3 \emph{(Additivity)} and F4 \emph{(Projectivity)}~\footnote{%
See \citep{Ma83}.}. This is done in one go,
for arbitrary (suitably typed) $R, f, g, h$~\footnote{In the Hoare logic counterpart of this rule
$gh$ will be the conjunction of two assertions.}:
\begin{eqnarray}
	\lfdep {f} {R} {g h} & \equiv & \arrayin{ \lfdep {f} {R} {g} & \land & \lfdep f R h }
	\label{eq:041106b}
\end{eqnarray}
Calculation:
\begin{eqnarray*}
&&
	\lfdep f R {g h}
\just\equiv{ (\ref{eq:050109b}) ; expansion of shorthand $g h$  }
	\join  g h \leq f · R°
\just\equiv{ universal property of $\Join$ (\ref{eq:050112a}) }
	g \leq f · R° \wider\land h \leq f · R°
\just\equiv{ (\ref{eq:050109b}) twice }
%
	\arrayin{
	\lfdep f R g
& \land &
	\lfdep f R  h
}
\end{eqnarray*}
The type rule for the database join operator ($\Join$) is calculated in the same way:
\begin{eqnarray*}
&&
	\lfdep{f}{R}{g} \land \lfdep{f}{S}{h}
\just\entails{ let $\p1(y,x)=y$ and $\p2(y,x)=x$; FDs are downward closed }  
	\longlfdep{f}{\p1· (R \Join S)}{g} \land \longlfdep{f}{\p2· (R \Join S)}{h}
\just\equiv{ trading (\ref{eq:061227a-modified}) twice }
	\lfdep{f}{{R \Join S}}{g· \p1} \land \lfdep{f}{{R \Join S}}{h· \p2}
\just\equiv{ F3+F4 (\ref{eq:041106b}) }
	\lfdep{f}{{R \Join S}}{(g· \p1)\Join(h· \p2)}
\just\equiv{ product of functions: $f\times g=(f\comp\p1)\Join(g\comp\p2)$ }
	\lfdep{f}{R \Join S}{g× h}
\end{eqnarray*}

\section{Beyond the type system: query optimization}
\label{sec:120312e}
As explained above, FD theory (cf.\ Hoare logic) can be regarded as a type system
whose rules help in reasoning about data models (cf.\ programs) without
going into the semantic intricacies of data business rules (cf.\ program meanings).
It helps because quantified expressions such as in Definition \ref{def:041216a}
don't scale up very well to large sets of dependencies. In this respect, our
quantifier-free equivalent  (\ref{eq:050109b}) looks more tractable and is
therefore expected to be calculationally effective where the quantified equivalent
is clumsy.

This will be illustrated below with a simple example, taken from \cite{AHV95}
and also addressed by \cite{Wi12}: one wants to optimize the conjunctive query
\begin{eqnarray}
	\{ (d, a') \ | \ t=t', (t,d,a) \in \Movies, (t',d',a') \in \Movies \}
	\label{eq:120309a}
\end{eqnarray}
over a database file $\Movies(Title, Director, Actor)$ into a query accessing this
file only once, knowing that 
FD $\rarrow{Title}{}{Director}$ holds.

Put in calculational format and abbreviating $M$ for $\Movies$,
$t$ for $Title$, $d$ for $Director$ and $a$ for $Actor$, we want
to solve for $X$ the equation
\begin{eqnarray}
	d· M· (\ker t) · M· a° = X
	\label{eq:091220a}
\end{eqnarray}
whose left hand side is the relational equivalent of (\ref{eq:120309a}) \footnote{
As the interested reader may check by introducing the variables back. Note
how $\ker t$ expresses $t=t'$ and projection functions $d$ (for $Director$)
and $a$ (for $Actor$) work over tuple $(t,d,a)$ and tuple $(t',d',a')$,
respectively. The use of the same letters for data variables and the corresponding
projection functions should help in tallying the two versions of the query.
}.
Our aim is to obtain a solution $X$ containing only one instance of $M$.
The equation is solved by taking the FD itself as starting point and trying to
re-write it into something one recognizes as an instance of (\ref{eq:091220a}):
\begin{eqnarray*}
&&
	\lfdep t M d
\just\equiv{ (\ref{eq:050109b}) }
	d \leq t · M°
\just\equiv{ expanding (\ref{eq:020624a},\ref{eq:041217a}); $\conv M=M$ since $M$ is a set }
	M \comp \conv t \comp t \comp M \subseteq \conv d \comp d
\just\equiv{ composition $(\comp M)$ with a set (partial identity) is a closure operator }
	M \comp \conv t \comp t \comp M \subseteq \conv d \comp d \comp M
\just\entails{ shunting (\ref{eq:020617e},\ref{eq:020617f}); monotonicity of $(· a°)$; kernel (\ref{eq:020624a}) }
	d \comp M \comp (\ker t) \comp M \comp \conv a \subseteq d \comp M \comp \conv a 
\end{eqnarray*}
Thus we find $d \comp M \comp \conv a$ as a candidate solution for $X$. To
obtain $X = d \comp M \comp \conv a$ it remains to check the converse inclusion:
\begin{eqnarray*}
&&
	d \comp M \comp \conv a \subseteq d \comp M \comp (\ker t) \comp M \comp \conv a 
\just\entailed{ $id \subseteq \ker t$ because kernels are equivalence relations }
	d \comp M \comp \conv a \subseteq d \comp M \comp M \comp \conv a 
\just\equiv{ $M \comp M = M \cap M = M$ because $M$ is a set }
	d \comp M \comp \conv a \subseteq d \comp M \comp \conv a 
\end{eqnarray*}
Thus $X=d \comp M \comp \conv a$, that is
\[
	X \wider= \{ (d, a') \ | \ (t,d,a') \in \Movies \}
\]
is the solution to equation (\ref{eq:091220a}) which optimizes
the given query by only visiting the movies file once \footnote{
By the way: symmetry between $a$ and $d$ in calculation step
\(
	d· M· t°· t· M· a° \subseteq d· M· a°
\)
above immediately tells that FD $\lfdep t M a$ would also enable the
proposed  optimization.}.

\section{Conclusions and future work}
~ \hfill
\begin{minipage}{.80\textwidth}\footnotesize\em
``The great merit of algebra is as a powerful tool for exploring family relationships over a wide
range of different theories.
(...)
It is only their algebraic properties that emphasise the family likenesses (...)
Algebraic proofs by term rewriting are the most promising way in which computers
can assist in the process of reliable design.''
\\ \em
\rule{0pt}{1.5em} \hfill \cite{HoJi98}
\\\rule{0pt}{1.5em}
\end{minipage}

\noindent
There is growing interest in applying abstract algebra techniques in computer
science as a way to promote calculation in software engineering. Moreover,
algebraic structures such as idempotent semirings and Kleene algebras (which
relation algebra is an instance of) have been shown to be amenable to automation
\citep{HS07}.  \cite{MRE12}, for instance, encode a database preference theory
into idempotent semiring algebra and show how to use Prover9 to discharge proofs.
Model checking in tools such as eg.\ the Alloy Analyser also blends well with
quantifier-free relational models \citep{OF12}.

Abstract algebra has the power to \emph{unify} seemingly disparate theories
once they are encoded into the same abstract terms. In the current paper we have
shown how a relational rendering of both Hoare logic and data dependency
theory purports one such unification, in spite of the former being an algorithmic theory
and the latter a data theory, as both algorithms and data structures unify into
binary relations.

Other such unifications could be devised. For instance,
\cite{MDG85} reason about while-loops $w = (\textbf{while}\ t \ \textbf{do}\ b)$
in terms of so-called \emph{strongest invariant functions}, where invariant functions
$f$, ordered by injectivity, are such that
\(
	f \comp \crflx t = f \comp b       \comp \crflx t
\) holds. A simple argument in relation algebra shows this equivalent to 
\(
	f \comp b \comp \crflx t \subseteq f
\), thus entailing FD $\fdep f {b \comp \crflx t} f$.

On a more practical register, our algebraic framework makes it possible to type-check database
operations and optimize queries by calculation once they are written as Tarskian,
quantifier-free formulas. We would like to investigate this further in connection to
\cite{Wi12}'s point-free query compiler.

Back to the opening story, surely Tarski's work on satisfaction and truth
is relevant to computer science. But Etchemendy's answer could have been
better tuned to the particular context of database technology
suggested by the Oracle towers landscape:
\begin{quote}
[...] \qemph{%
They would never have been built without Tarski's work on the
{calculus of binary relations}.}
\end{quote}

\paragraph{Acknowledgments.}
This work is funded by ERDF - European Regional Development Fund through
the COMPETE Programme (operational programme for competitiveness) and by
National Funds through the \emph{FCT - Funda\c{c}\~ao para a Ci\^encia e a Tecnologia}
(Portuguese Foundation for Science and Technology) within project FCOMP-01-0124-FEDER-010047.

Feedback and exchange of ideas with Ryan Wisnesky about pointfree query reasoning are
gratefully acknowledged.
The author would also like to thank the anonymous reviewers whose comments helped to
improve the manuscript.


\end{document}